\documentclass[aip,jmp,amsmath,amssymb,preprint]{revtex4-1}

\usepackage{natbib}

\usepackage{amsmath}

\usepackage{amssymb}

\usepackage{bm}

\begin{document}

\bibliographystyle{plainnat}

\title{Derivatives of the Pochhammer and reciprocal Pochhammer symbols and their use in epsilon-expansions of Appell and Kamp\'e de F\'eriet functions}

\author{David Greynat}
\email{david.greynat@gmail.com}
\affiliation{INFN Sezione di Napoli, Dipartimento di Fisica,\\Via Cintia, I-80126 Napoli, Italy}

\author{Javier Sesma}
\email{javier@unizar.es}
\affiliation{Departamento de F\'{\i}sica Te\'orica, Facultad de Ciencias,\\ 50009, Zaragoza, Spain}

\author{Gr\'egory Vulvert}
\email{Gregory.Vulvert@ific.uv.es}
\affiliation{Departament de F\'{\i}sica Te\'orica, IFIC, CSIC - Universitat de Val\`{e}ncia,\\ Apt. Correus 22085,  E-46071 Val\`{e}ncia, Spain}

\date{\today}

\begin{abstract}
Useful expressions of the derivatives, to any order, of Pochhammer and reciprocal Pochhammer symbols with respect to their arguments are presented. They are building blocks of a procedure, recently suggested, for obtaining the $\varepsilon$-expansion of functions of the hypergeometric class related to Feynman integrals. The procedure is applied to some examples of such kind of functions taken from the literature.
\end{abstract}



\maketitle

\section{Introduction}

Calculations in Quantum Field theories are based on series expansions in the coupling
constants involved in their Lagrangians. Each coefficient of those series is a
sum of Feynman diagrams which constitute the main tools for producing physical predictions.
 To cure the fact that some of these diagrams involve divergent integrals in the
four dimensional space-time, a dimensional regularization is widely used. Basically, it consists in the analytic continuation of integrals from dimension D=4 to an arbitrary D-dimensional space-time where all integrals become convergent. A
conventional way of doing it is to set D$=\! 4\! -\! \varepsilon$ and to perform the calculations as expansions in powers of $\varepsilon$.
On the other hand, one can always write a Feynman diagram through its
Feynman parametrization as
\[
\mathcal{F}({\bm{\rho}},\mathrm{D}\! =\! 4\! -\! 2\varepsilon)=\int_0^1dx_1 \ldots\int_0^1dx_N\,\frac{\mathcal{N}({\mathbf{x}})}{\left[\sum_{n\geq 1} \mathcal{D}_n(\mathbf{x})\,\rho_n\right]^{r+\frac{p}{q}\varepsilon}}\,
\]
where $\rho_n$ are the scalar products of the external momenta and squared masses in the diagram, $\mathcal{N}$
and $\mathcal{D}_n$ are real polynomials of the $N$ Feynman parameters, and $p$, $q$, $r$ are integers related to the
numbers of loops, external momenta and topology of the diagrams. It appears that, through this
formulation, there is a strong link between Feynman diagrams and generalized
hypergeometric functions whose parameters depend linearly on $\varepsilon$, as shown in the denominator of the integrand in the right hand side of the equation above. Then, calculations of Feynman diagrams benefit from the so called $\varepsilon$-expansion of those functions of the hypergeometric class, that is, their series expansion in powers of $\varepsilon$.

A considerable number of papers dealing with the $\varepsilon$-expansion of generalized hypergeometric functions have appeared in the last decade. Besides analytical methods \cite{jege,dav4,kalm,kal1,kal2,kal3,yost} applicable for restricted values of the parameters, computer packages \cite{moc1,wei1,wei2,moc2,hub1,hub2,huan,byte} based on different algorithms have been published.

In a recent paper \cite{grey} we suggested a simple procedure to evaluate the successive terms of the $\varepsilon$-expansion of functions of the hypergeometric class with the only restriction on the parameters of being linearly dependent on $\varepsilon$. Starting with the series expansions of those functions, we carried out a Taylor expansion around the value $\varepsilon=0$. Essential elements in the proposed algorithm are derivatives, to any order, of the Pochhammer and reciprocal Pochhammer symbols with respect to their arguments. The procedure was illustrated by
applying it to get the $\varepsilon$-expansion of generalized hypergeometric functions of the type $\ _{q+1}\!F_q$.

Ancarani and Gasaneo, in a series of papers,\cite{anca}  have addressed the problem of finding expressions for the derivatives, of arbitrary order, of the hypergeometric function $\ _p\! F_q$ with respect to one of its (upper or lower) parameters. By making use of Babister's solution to nonhomogeneous differential equations satisfied by the derivatives of $\ _p\! F_q$, they have written those derivatives in terms of multivariable Kamp\'e de F\'eriet functions. The resulting expressions are certainly beautiful, but we do not find them applicable to the $\varepsilon$-expansion problem.

The purpose of this paper is twofold. Firstly, in order to get the maximum efficiency in the computation of $\varepsilon$-expansions, we present, in Sections II and III, a collection of properties of derivatives of the Pochhammer and reciprocal Pochhammer symbols, respectively, and equations relating them to other mathematical objects. With the same goal, we suggest in Section IV a partial fraction decomposition of the quotient of two Pochhammer symbols, previous to derivation with respect to $\varepsilon$. The second part of the paper is devoted to calculate, by benefiting from the results obtained in the previous sections, the $\varepsilon$-expansion of some examples of functions of the hypergeometric class, namely Appell and Kamp\'e de F\'eriet functions, related to Feynman integrals. In particular, we obtain, in Section V, all terms of the $\varepsilon$-expansion of the Appell functions appearing in a paper by Del Duca  {\it et al.},\cite{deld} dealing with the one-loop pentagon, and in another paper by Tarasov,\cite{tara} concerned with the two loop equal mass sunrise diagram. Kamp\'e de F\'eriet functions appearing in Ref. 18 are considered in section VI. Partial fraction decomposition of quotients of several Pochhammer symbols and its usefulness in the derivation  of $\varepsilon$-expansions are discussed in Section VII.

Along the paper we use the notation
\begin{equation}
\alpha_i \equiv A_i+a_i\,\varepsilon\,,\qquad \beta_j \equiv B_j+b_j\,\varepsilon\,,  \label{i1}
\end{equation}
for the upper and lower parameters of the functions of the hypergeometric class related to the Feynman integrals. (Notice the linear dependence of $\alpha$ and $\beta$ on $\varepsilon$.) Essential elements of our procedure to obtain the $\varepsilon$- expansion of those functions are the derivatives, to any order, of the Pochhammer and reciprocal Pochhammer symbols with respect to their argument, which we denote by
\begin{equation}
\mathcal{P}_m^{(k)}(\alpha) \equiv \frac{1}{k!}\,\frac{d^k}{d\alpha^k}(\alpha)_m\,,  \qquad
\mathcal{Q}_{m}^{(k)}(\beta) \equiv \frac{1}{k!}\,\frac{d^k}{d\beta^k}\frac{1}{(\beta)_m}\,.  \label{i3}
\end{equation}
Besides those essential elements, we will use the successive derivatives with respect to $\varepsilon$ of the quotient of two Pochhammer symbols, which we denote by
\begin{equation}
\mathcal{R}_{m,n}^{(k)}(\alpha, \beta) \equiv \frac{1}{k!}\,\frac{\partial^k}{\partial\varepsilon^k}\frac{(A+a\,\varepsilon)_m}{(B+b\,\varepsilon)_n}\,. \label{i2}
\end{equation}

\section{Derivatives of the Pochhammer symbol}

As $(\alpha)_m$ is a polynomial of degree $m$ in $\alpha$, its derivatives with respect to $\alpha$ of order larger than $m$ vanish, that is, with the notation introduced in Eq. (\ref{i3}),
\begin{equation}
\mathcal{P}_m^{(k)}(\alpha) = 0 \qquad\mbox{for}\qquad k>m .  \label{a2}
\end{equation}
We consider in this section the non-trivial case of being $k\leq m$. Obviously, the successive derivatives $\mathcal{P}_m^{(k)}(\alpha)$ can be computed by means of the recurrence relation
\begin{equation}
\mathcal{P}_{m+1}^{(k)}(\alpha) = (\alpha+m)\,\mathcal{P}_m^{(k)}(\alpha)+\mathcal{P}_m^{(k-1)}(\alpha), \qquad k>0, \label{a10}
\end{equation}
with starting values
\begin{equation}
\mathcal{P}_{0}^{(k)}(\alpha) = \delta_{k,0}, \qquad \mathcal{P}_m^{(0)}(\alpha)=(\alpha)_m, \label{a11}
\end{equation}
Explicit expressions of the $\mathcal{P}_m^{(k)}$ can be obtained directly from a generating relation which can be immediately deduced from
\begin{equation}
\sum_{m=0}^\infty (\alpha)_m (-t)^m/m! \equiv \ _1\!F_0(\alpha;;-t) = (1+t)^{-\alpha}, \qquad |t|<1  \label{a3}
\end{equation}
(see Eq. (2) in Section 6.2.1 of Ref. 20). Derivation, $k$ times, with respect to $\alpha$ gives
\begin{equation}
\sum_{m=0}^\infty k!\, \mathcal{P}_m^{(k)}(\alpha)\, (-t)^m/m! = (-1)^k\,(1+t)^{-\alpha}\, \left[\ln(1+t)\right]^k, \qquad |t|<1. \label{a4}
\end{equation}
Then, we have
\begin{eqnarray}
\hspace{-1cm}\mathcal{P}_m^{(k)}(\alpha) &=& \frac{(-1)^{k-m}}{k!}\,\left.\frac{\partial^m}{\partial t^m}\left((1+t)^{-\alpha}\,
\left[\ln(1+t)\right]^k\right)\right|_{t=0}  \nonumber \\
&=& \frac{(-1)^{k-m}}{k!}\,\sum_{l=0}^m {m \choose l}\,\left.\left(\frac{\partial^l}{\partial t^l}(1+t)^{-\alpha}\right)\left(\frac{d^{m-l}}{d
t^{m-l}}\left[\ln(1+t)\right]^k\right)\right|_{t=0}\hspace{-12pt}. \label{a5}
\end{eqnarray}
Obviously,
\begin{equation}
\left.\frac{\partial^l}{\partial t^l}(1+t)^{-\alpha}\right|_{t=0} = (-1)^l\,(\alpha)_l.  \label{a6}
\end{equation}
On the other hand, since $[\ln (1+t)]^k$ is the generating function of the Stirling numbers of the first kind,
\begin{equation}
[\ln (1+t)]^k = k! \sum_{n=k}^\infty s(n, k)\, t^n/n!\,, \qquad |t|<1  \label{a7}
\end{equation}
(see Eq. 26.8.8 of Ref. 21), we get
\begin{equation}
\left.\left(\frac{d^{m-l}}{d t^{m-l}}[\ln(1+t)]^k\right)\right|_{t=0} = k!\,s(m-l, k)\,.  \label{a8}
\end{equation}
Substitution of Eqs. (\ref{a6}) and (\ref{a8}) in (\ref{a5}) gives
\begin{equation}
\mathcal{P}_m^{(k)}(\alpha) = (-1)^{m-k}\sum_{l=0}^{m-k}(-1)^l{m \choose l}\,s(m-l, k)\,(\alpha)_l \qquad\mbox{for}\quad m\geq k. \label{a9}
\end{equation}

Coffey, in two papers \cite{cof1,cof2} dealing with expansions of the Hurwitz zeta function, has given two useful alternative expressions of the derivatives of the Pochhammer symbol. The first one (see Eq. (2.3) of Ref. 22), after correcting an obvious misprint, reads
\begin{equation}
\left(\frac{d}{ds}\right)^l\,(s)_j=\sum_{k=l}^j(-1)^{j+k}\,s(j,k)\,k(k-1)\cdots(k-l+1)\,s^{k-l}, \label{coffey1}
\end{equation}
that is, in our notation,
\begin{equation}
\mathcal{P}_m^{(k)}(\alpha) = (-1)^{m-k}\sum_{j=0}^{m-k}(-1)^j{k\! +\! j \choose k}s(m,k\! +\! j)\,\alpha^j, \quad \mbox{for}\; m\geq k\,. \label{A7}
\end{equation}
The second alternative form of $\mathcal{P}_m^{(k)}(\alpha)$ stems from the expression (see Eq. (2.5) of Ref. 23)
\begin{equation}
B_\nu^{(n+1)}(x)=(-1)^n\,\frac{\nu!}{n!}\,\frac{d^{n-\nu}}{dx^{n-\nu}}(1-x)_n \,, \qquad \nu\leq n\,, \label{A17}
\end{equation}
for the generalized Bernoulli polynomials. These  are generated by the relation \cite{sriv,bryc}
\begin{equation}
\left(\frac{z}{e^z-1}\right)^a\,e^{xz}=\sum_{n=0}^\infty B_n^{(a)}(x)\,\frac{z^n}{n!}\,, \qquad |z|<2\pi\,. \label{A18}
\end{equation}
(Notice that the value $a=1$ corresponds to the familiar Bernoulli polynomials.)
It is immediate to obtain from Eq. (\ref{A17}) the useful relation
\begin{equation}
\mathcal{P}_m^{(k)}(\alpha) = (-1)^{m-k}\,{m \choose k}\,B_{m-k}^{(m+1)}(1\! -\! \alpha)\,.  \label{A19}
\end{equation}

The explicit expressions of the successive derivatives of the Pochhammer  symbol given in Eqs. (\ref{a9}), (\ref{A7}) and (\ref{A19}) become considerably simplified for certain particular values of the argument $\alpha$. For $\alpha=0$ one has
\begin{eqnarray}
\mathcal{P}_0^{(k)}(0)&=& \delta_{k,0},  \qquad \mathcal{P}_m^{(0)}(0)= \delta_{m,0}, \qquad  \mathcal{P}_m^{(k)}(0)=0 \quad\mbox{for}\quad k>m\,,
\label{A5}  \\
\mathcal{P}_m^{(k)}(0)&=& (-1)^{m-k}\,s(m, k) \qquad \mbox{for}\quad m\geq k>0\,. \label{A6}
\end{eqnarray}
These expressions allow one to write the MacLaurin expansion of $\mathcal{P}_m^{(k)}(\alpha)$ and to reproduce, in this way, the expression (\ref{A7}) due to Coffey.
Another interesting particular value of the argument of $\mathcal{P}_m^{(k)}$ is 1. In view of the relation
\begin{equation}
s(n+1,k+1)=n!\sum_{j=k}^n \frac{(-1)^{n-j}}{j!}\,s(j, k)   \label{A8}
\end{equation}
(see Eq. 26.8.20 in Ref. 21), one obtains from Eq. (\ref{a9})
\begin{equation}
\mathcal{P}_m^{(k)}(1) = (-1)^{m-k}\,s(m+1, k+1)\,,  \label{A9}
\end{equation}
in terms of Stirling numbers of the first kind. This result, as a lemma, was already proved by Coffey (see Eq. (2.1) of Ref. 22). Alternatively, one can deduce from Eq. (\ref{A19}) the representation
\begin{equation}
\mathcal{P}_m^{(k)}(1) = (-1)^{m-k}\,{m \choose k}\,B_{m-k}^{(m+1)}\,  \label{AA19}
\end{equation}
in terms of generalized Bernoulli numbers.
Another expression of $\mathcal{P}_m^{(k)}(1)$, as an Euler-Zagier sum,\cite{moc1,verm}
\begin{equation}
\mathcal{P}_m^{(k)}(1) = \left.\frac{1}{k!}\,\frac{d^k}{d\varepsilon^k}(1+\varepsilon)_m \right|_{\varepsilon =0} = m!\,Z_{1,\ldots,1}(m) \quad (k \;  \mathrm{subindices}), \label{A15}
\end{equation}
can be found in the paper by Del Duca {\em et al.} (see Eq. (5.17) of Ref. 18).
From Eq. (\ref{A9}), it is immediate to write the Taylor expansion
\begin{equation}
\mathcal{P}_m^{(k)}(\alpha) = \sum_{j=k}^{m}(-1)^{m-j}{j \choose k}\,s(m\! +\! 1, j\! +\! 1)\,(\alpha\! -\! 1)^{j-k}\,, \quad\mbox{for}\quad m\geq k\,.  \label{A13}
\end{equation}
An alternative expression valid for integer $\alpha\neq 0, 1$ will be given in the next section.

\section{Derivatives of the reciprocal Pochhammer symbol}

There is an obvious recurrence relation which allows one to compute the derivatives, to any order, of the reciprocal Pochhammer symbol $1/(\beta)$ with respect to $\beta$. It is
\begin{equation}
\mathcal{Q}_{m+1}^{(k)}(\beta) = \left(\mathcal{Q}_{m}^{(k)}(\beta) - \mathcal{Q}_{m+1}^{(k-1)}(\beta)\right)/(\beta+m),  \qquad k>0, \label{b8}
\end{equation}
with starting values
\begin{equation}
\mathcal{Q}_0^{(k)}(\beta) = \delta_{k,0}, \qquad   \mathcal{Q}_{m}^{(0)}(\beta) = 1/(\beta)_m. \label{b9}
\end{equation}

Very simple expressions for the $\mathcal{Q}_{m}^{(k)}(\beta)$ can be easily obtained from the relation
\begin{equation}
\frac{1}{(\beta)_m} = \sum_{l=0}^{m-1}\frac{(-1)^l}{l!\,(m-1-l)!}\,\frac{1}{\beta+l}\,, \qquad m>0   \label{b2}
\end{equation}
(see Eq. 4.2.2.45 in Ref. 27). Direct derivation with respect to $\beta$ in this equation gives
\begin{equation}
\mathcal{Q}_{m}^{(k)}(\beta)= (-1)^k \sum_{l=0}^{m-1}\frac{(-1)^l}{l!\,(m-1-l)!}\,\frac{1}{(\beta+l)^{k+1}}\,, \qquad m>0\,,  \label{b3}
\end{equation}
provided that $\beta$ is different from a non-positive integer, $-n$, such that $0\leq n<m$. In the case of one of the lower parameters of the function to be expanded being of the form
\[
\beta=-n+b\,\varepsilon,   \qquad n\geq 0\,,
\]
one may use, for $m>n$, the evident relation
\[
\frac{1}{(-n+b\,\varepsilon)_m}=\frac{1}{b\,\varepsilon}\,\frac{(-1)^n}{(1-b\,\varepsilon)_n}\,\frac{1}{(1+b\,\varepsilon)_{m-n-1}}\,.
\]
The factor $1/b\varepsilon$ should be isolated and the Taylor expansion would affect only the other factors. The final $\varepsilon$-expansion could then contain negative powers of $\varepsilon$.

A concise form of Eq. (\ref{b3}), in terms of the difference operator $\Delta$, can be written immediately by comparison of the right hand side of that equation and that of the well known property (see Ref. 28, Sec. 1.6, Eq. [6c], or Ref. 29, Sec. 6.2, Eq. (1))
\begin{equation}
\Delta^n f(x)=\sum_{k=0}^n(-1)^{n-k}\,{n \choose k}\,f(x+k)\,.   \label{comtet1}
\end{equation}
The result is
\begin{equation}
\mathcal{Q}_{m}^{(k)}(\beta)= \frac{(-1)^{m-k-1}}{(m-1)!}\,\Delta^{m-1}\frac{1}{\beta^{k+1}}\,, \qquad m>0\,,  \label{comtet2}
\end{equation}

For the particular case of $\beta=1$, the derivatives of the reciprocal Pochhammer symbol admit much simpler forms.
In analogy with the generalized harmonic numbers,
\begin{equation}
H_m^{(k)} \equiv \sum_{j=1}^m \frac{1}{j^k}\,, \qquad m\geq 1\,,  \label{A10}
\end{equation}
we define modified generalized harmonic numbers
\begin{equation}
\hat{H}_m^{(k)} \equiv \sum_{j=1}^m (-1)^{j-1}\,{m \choose j}\,\frac{1}{j^k}\,, \qquad m\geq 1\,,   \label{A11}
\end{equation}
whose sequence, $\{\hat{H}_n^{(k)}\}$, $n=1, 2, 3, \ldots$ turns out to be the conjugate, in the sense defined by Vermaseren (see Appendix B of Ref. 26) and used by Moch {\em et al.} (see Eq. (29) of Ref. 8), of the sequence $\{1/n^k\}$. From Eqs. (\ref{b3}) and (\ref{A11}), it is immediate to write
\begin{equation}
\mathcal{Q}_m^{(k)}(1)=\frac{(-1)^k}{m!}\,\hat{H}_m^{(k)}\,, \qquad m\geq 1\,.  \label{A12}
\end{equation}
This relation can be extended to $m=0$ with the convention
\begin{equation}
\hat{H}_0^{(k)}=\delta_{k,0}\,. \label{Ah}
\end{equation}
For values of $\beta$ different from 1, it is immediate to write the Taylor expansion
\begin{equation}
\mathcal{Q}_m^{(k)}(\beta) = \frac{1}{m!}\,\sum_{j=0}^\infty (-1)^{k+j}{k+j \choose k}\,\hat{H}_m^{(k+j)}\,(\beta-1)^j\,, \qquad m\geq 1\,.  \label{A14}
\end{equation}
 Del Duca {\em et al.} have given, for $\mathcal{Q}_m^{(k)}(1)$, an alternative expression,
\begin{equation}
\mathcal{Q}_m^{(k)}(1) =  \left.\frac{1}{k!}\,\frac{d^k}{d\varepsilon^k}\frac{1}{(1+\varepsilon)_m} \right|_{\varepsilon =0} = - \, \frac{1}{m!}\,S_{1,\ldots,1}(m) \quad (k\; \mbox{subindices}),   \label{A16}
\end{equation}
(see Eq. 5.17 of Ref. 18) in terms of nested harmonic sums.\cite{moc1,verm}

For given values of $k$ and $\beta$, a generating function of the sequence $\{\mathcal{Q}_{n+1}^{(k)}(\beta)\}$, $n=0, 1, 2, \dots$, can be easily found. Let us recall the binomial transform,\cite{goul,flaj,hauk,prod} $T$, relating two sequences $\{a_n\}$ and $\{b_n\}=T\{a_n\}$, $n=0, 1, 2, \dots$, in this way:
\begin{equation}
b_n=\sum_{l=0}^n {n \choose l}\,a_l\,,\qquad a_n=\sum_{j=0}^n (-1)^{n-j}\,{n \choose j}\,b_j\,. \label{A20}
\end{equation}
The ordinary generating functions of those sequences,
\begin{equation}
f(z)=\sum_{n=0}^\infty a_n\,z^n \quad \mbox{and} \quad g(z)=\sum_{n=0}^\infty b_n\,z^n\,,   \label{A21}
\end{equation}
are related by the Euler transform
\begin{equation}
g(z)=\frac{1}{1-z}\,f\left(\frac{z}{1-z}\right)\,.  \label{A22}
\end{equation}
From Eq. (\ref{b3}), written in the form
\begin{equation}
(-1)^k\,m!\,\mathcal{Q}_{m+1}^{(k)}(\beta) =  \sum_{l=0}^{m} {m \choose l}\,\frac{(-1)^l}{(\beta+l)^{k+1}}\,,  \label{A23}
\end{equation}
one realizes that the sequence $\{(-1)^k\, n!\,\mathcal{Q}_{n+1}^{(k)}(\beta)\}$, $n=0, 1, 2, \ldots$, is the binomial transformed of the sequence $\{(-1)^n/(\beta+n)^{k+1}\}$. Since the ordinary generating function of this sequence is a Lerch function (see \S25.14 of Ref. 21), namely
\begin{equation}
\Phi(-z,k+1,\beta)=\sum_{n=0}^\infty \frac{(-z)^n}{(\beta+n)^{k+1}}\,, \qquad |z|<1\,, \label{A24}
\end{equation}
one can write immediately the generating relation
\begin{equation}
\frac{1}{1-z}\,\Phi\left(\frac{-z}{1-z},k+1,\beta\right)=\sum_{m=0}^\infty (-1)^k\,m!\,\mathcal{Q}_{m+1}^{(k)}(\beta)\,z^m  \label{A25}
\end{equation}
for the derivatives of the reciprocal Pochhammer symbols.

Interesting properties of the modified generalized harmonic numbers, defined in Eq. (\ref{A11}), can be easily obtained. From Eq. (\ref{A25}), by taking $\beta=1$, one gets the generating relation, for fixed $k$,
\begin{equation}
-\,{\rm Li}_{k+1}\left(\frac{-z}{1-z}\right) = \sum_{m=1}^\infty \frac {1}{m}\,\hat{H}_m^{(k)}\,z^m\,,   \label{A26}
\end{equation}
where Li$_n$ represents the polylogarithm function. Another generating relation, for fixed $m\geq 1$, reads
\begin{equation}
\prod_{j=1}^m\left(\frac{1}{1-(c_m/j)z}\right) = \sum_{k=0}^\infty \,c_m^k\, \hat{H}_m^{(k)}\,z^k\,,  \qquad |z|<1/c_m\,,  \label{nueva1}
\end{equation}
where the coefficient $c_m$ may be chosen arbitrarily. Possibly interesting choices seem to be
\[
c_m=1\,, \qquad c_m={\rm l.c.m.}(1, 2, \ldots, m)\,, \qquad c_m=m!\,.
\]
Equation (\ref{nueva1}) can be checked by decomposing its left hand side in partial fractions,
\begin{equation}
\prod_{j=1}^m\left(\frac{1}{1-(c_m/j)z}\right) = \sum_{j=1}^m (-1)^{j-1}{m \choose j}\frac{1}{1-(c_m/j)z}\,,  \label{nueva2}
\end{equation}
by using the geometric series expansion
\begin{equation}
\frac{1}{1-(c_m/j)z} = \sum_{k=0}^\infty (c_m/j)^k\,z^k\,, \qquad {\rm provided} \quad |z|<1/c_m\,, \label{nueva3}
\end{equation}
and by recalling the definition (\ref{A11}) of $\hat{H}_m^{(k)}$.

From the evident relation
\begin{equation}
\sum_{j=0}^k\,\mathcal{P}_m^{(k-j)}(x)\,\mathcal{Q}_m^{(j)}(x)= \delta_{k,0}\,,  \label{A27}
\end{equation}
one obtains, in the case of $x=1$, the sum rule
\begin{equation}
\sum_{j=0}^k\,s(m\! +\! 1,k\! +\! 1\! -\! j)\,\hat{H}_m^{(j)}= \delta_{k,0}\,(-1)^{m}\,m!\,,  \label{A28}
\end{equation}

Combination of Eqs. (\ref{A9}) and (\ref{A12}) allows one to write, for the cases of integer argument $n\geq 1$,
\begin{eqnarray}
\mathcal{P}_m^{(k)}(n) &=& \frac{(-1)^{m+n-1-k}}{(n-1)!}\,\sum_{j=0}^{k}\,s(m\! +\! n, k\! +\! 1\! - \! j)\,\hat{H}_{n-1}^{(j)}\,, \label{A29}  \\
\mathcal{Q}_m^{(k)}(n) &=& \frac{(-1)^{n-1-k}}{(m+n-1)!}\,\sum_{j=0}^{k}\,s(n, k\! +\! 1\! - \! j)\,\hat{H}_{m+n-1}^{(j)}\,. \qquad  \label{A30}
\end{eqnarray}
In the case of negative integer argument one obtains
\begin{eqnarray}
\mathcal{P}_m^{(k)}(-n) &=& \left\{\begin{array}{ll}(-1)^{m-k} \,\mathcal{P}_m^{(k)}(n\! +\! 1\! -\! m) \qquad {\rm for}\; m\leq n \,,\\  (-1)^{m-n-k}\sum_{j=0}^{k-1}\,(-1)^j\,s(n\! +\! 1,j\! +\! 1)\,s(\! m\! -\! n,k\! -\! j) \qquad  {\rm for}\; m\geq n\! +\! 1 \,, \end{array}\right.  \label{A31}  \\
\mathcal{Q}_m^{(k)}(-n) &=& (-1)^{m-k}\,\mathcal{Q}_m^{(k)}(n+1-m)\qquad {\rm for}\; m\leq n \,.  \label{A32}
\end{eqnarray}

\section{Partial fraction decomposition of the quotient of two Pochhammer symbols}

The multiple derivation with respect to $\varepsilon$, inherent to our procedure for obtaining the $\varepsilon$-expansion, is considerably simplified by a previous decomposition of the quotient of two Pochhammer symbols in partial fractions.
The basic equation for this purpose can be found in the Digital Library of Mathematical Functions (see Eq. 1.2.12 of Ref. 21). It reads
\begin{equation}
\frac{f(x)}{(x-x_1)(x-x_2)\cdots (x-x_n)} = \frac{C_1}{x-x_1}+\frac{C_2}{x-x_2}+ \dots +\frac{C_n}{x-x_n}, \label{ii1}
\end{equation}
where $f(x)$ represents a polynomial  of degree less than $n$ and the coefficients $C_j$ are given by
\begin{equation}
C_j = \frac{f(x_j)}{\prod_{k\neq j}(x_j-x_k)}.  \label{ii2}
\end{equation}
Equation (\ref{ii1}) can be extended trivially to the case of $f(x)$ being a polynomial $p_n(x)$ of degree $n$,
\begin{equation}
p_n(x)=c_0+c_1\,x+\ldots +c_n\,x^n\,.   \label{ii3}
\end{equation}
The result is
\begin{equation}
\frac{p_n(x)}{(x-x_1)(x-x_2)\cdots (x-x_n)} =c_n + \frac{C_1}{x\! -\! x_1}+\frac{C_2}{x\! -\! x_2}+ \dots +\frac{C_n}{x\! -\! x_n}, \label{ii4}
\end{equation}
the coefficients $C_j$ being now given by
\begin{equation}
C_j = \frac{p_n(x_j)}{\prod_{k\neq j}(x_j-x_k)}.  \label{ii5}
\end{equation}

Bearing in mind those partial fraction expansions, it is immediate to obtain
\begin{eqnarray}
\mathcal{R}_{m,n}^{(0)}(\alpha, \beta)&\equiv& \frac{(A+a\,\varepsilon)_m}{(B+b\,\varepsilon)_n} = \sum_{j=0}^{n-1} \frac{r_{j,m,n}(\alpha,\beta)}{B+j+b\,\varepsilon},  \qquad \mbox{for} \quad m<n\,,       \label{ii9}  \\
\mathcal{R}_{n,n}^{(0)}(\alpha, \beta)&\equiv& \frac{(A+a\,\varepsilon)_n}{(B+b\,\varepsilon)_n} = \left(\frac{a}{b}\right)^n+\sum_{j=0}^{n-1} \frac{r_{j,n,n}(\alpha,\beta)}{B+j+b\,\varepsilon},          \label{ii10}
\end{eqnarray}
where
\begin{equation}
r_{j,m,n}(\alpha,\beta)=(-1)^j\,\frac{\Big(A-(a/b)(B+j)\Big)_m}{j!\,(n-1-j)!}\,,\qquad  \mbox{for} \quad m\leq n\,.       \label{ii11}  \\
\end{equation}
Derivation with respect to $\varepsilon$ in Eqs. (\ref{ii9}) and (\ref{ii10}) gives
\begin{eqnarray}
\mathcal{R}_{m,n}^{(k)}(\alpha, \beta)&=& (-b)^k\,\sum_{j=0}^{n-1} \frac{r_{j,m,n}(\alpha,\beta)}{(B+j+b\,\varepsilon)^{k+1}},
\qquad \mbox{for} \quad m<n\,,  \label{ii13}  \\
\mathcal{R}_{n,n}^{(k)}(\alpha, \beta)&=& (-b)^k\left(\delta_{k,0}\,\left(\frac{a}{b}\right)^n+\sum_{j=0}^{n-1} \frac{r_{j,n,n}(\alpha,\beta)}{(B+j+b\,\varepsilon)^{k+1}}\right),          \label{ii14}
\end{eqnarray}
For the case of being $m>n$, one may use one of the identities
\begin{equation}
\frac{(\alpha)_m}{(\beta)_n} = (\alpha)_{m-n}\,\frac{(\alpha+m-n)_n}{(\beta)_n}\,, \quad \mbox{or} \quad \frac{(\alpha)_m}{(\beta)_n}  =
\frac{(\alpha)_n}{(\beta)_n}\,(\alpha+n)_{m-n}\,,   \label{ii15}
\end{equation}
before deriving with respect to $\varepsilon$.

As a by-product of Eqs. (\ref{ii9}) and (\ref{ii10}), one can write the identities
\begin{eqnarray}
\frac{(A)_m}{(B)_n}&=&\sum_{j=0}^{n-1}\frac{(-1)^j\,\big(A-(B+j)\,x\big)_m}{j!\,(n-1-j)!\,(B+j)},\qquad \mbox{for}\quad m<n\,,  \label{ii16} \\
\frac{(A)_n}{(B)_n}&=& x^n+\sum_{j=0}^{n-1}\frac{(-1)^j\,\big(A-(B+j)\,x\big)_n}{j!\,(n-1-j)!\,(B+j)}, \label{ii17}
\end{eqnarray}
for arbitrary $A$, $B$, and $x$.

\section{$\varepsilon$-expansion of Appell functions}

The expressions obtained in the preceding sections are easily incorporated to the procedure for obtaining the $\varepsilon$-expansion of functions of the hypergeometric class. To illustrate how it can be done, we are going to obtain the $\varepsilon$-expansion of some Appell functions found in the literature. The first four ones are taken from a paper by Del Duca {\em et al.} (see Eq. (5.13) of Ref. 18) and the fifth one comes from a paper by Tarasov (see Eq. (4.32) of Ref. 19).

\subsection{1st example}

Let us start by considering
\begin{equation}
\mathcal{F}_1 \equiv F_4 \left( \left.\begin{array}{l}1-2\varepsilon,\,1-\varepsilon\\1-\varepsilon,\,1-\varepsilon\end{array}\right|x_1,\,x_2\right).  \label{iii1}
\end{equation}
We have for its series expansion
\begin{eqnarray}
\mathcal{F}_1 &=&
\sum_{m_1=0}^\infty \sum_{m_2=0}^\infty \frac{(1-2\varepsilon)_{m_1+m_2}\,(1-\varepsilon)_{m_1+m_2}}{(1-\varepsilon)_{m_1}\,(1-\varepsilon)_{m_2}}\,
\frac{x_1^{m_1}}{m_1!}\,\frac{x_2^{m_2}}{m_2!}  \nonumber  \\
&=&\sum_{m=0}^\infty \,\sum_{n=0}^m\, (1-2\varepsilon)_m\,\frac{(1+m-n-\varepsilon)_n}{(1-\varepsilon)_n}\,\frac{x_1^n}{n!}\,\frac{x_2^{m-n}}{(m-n)!}\,.
 \label{iii2}
\end{eqnarray}
The successive derivatives with respect to $\varepsilon$ can be written immediately in terms of the $\mathcal{P}_m^{(k)}(1-2\varepsilon)$ and the $\mathcal{R}_{n,n}^{(k)}(1+m-n-\varepsilon,\,1-\varepsilon)$. By using the explicit expression of $\mathcal{P}_m^{(k)}(1)$ given in Eq. (\ref{A9}), and that of $\mathcal{R}_{n,n}^{(k)}(1+m-n-\varepsilon,\,1-\varepsilon)|_{\varepsilon=0}$ resulting from (\ref{ii14}), one obtains the $\varepsilon$-expansion
\begin{eqnarray}
\mathcal{F}_1 &=& \sum_{k=0}^\infty \,\varepsilon^k \,\sum_{k_1=0}^k\,2^{k_1}\sum_{m=k_1}^\infty (-1)^m\,s(m+1, k_1+1)  \nonumber  \\
& & \times\,\sum_{n=0}^m\,\left(\delta_{k_1,k} - \sum_{j=1}^n(-1)^{j}{m-j \choose n}{n \choose j}\frac{1}{j^{k-k_1}}\right)\frac{x_1^n}{n!}\,\frac{x_2^{m-n}}{(m-n)!}\,.  \label{iii3}
\end{eqnarray}
Bearing in mind that
\begin{equation}
s(m+1, 1)=(-1)^m\,m!   \label{iii4}
\end{equation}
and (see Eq. 56 of Sec. 4.2.5 in Ref. 27)
\begin{equation}
1 - \sum_{j=1}^n(-1)^{j}{m-j \choose n}{n \choose j} = {m \choose n},  \label{iii5}
\end{equation}
it is easy to check that the apparently cumbersome expression of the coefficient of $\varepsilon^0$ reduces to
\begin{equation}
\sum_{m=0}^\infty\,\sum_{n=0}^\infty {m \choose n}^2\,x_1^n\,x_2^{m-n} = F_4 \left( \left.\begin{array}{l}1,\,1\\1,\,1\end{array}\right|x_1,\,x_2\right)\,.   \label{iii6}
\end{equation}
Besides, it is not difficult to see that the right hand side of Eq. (\ref{iii3}) is invariant in the interchange $x_1\Longleftrightarrow x_2$, as it should be.

\subsection{2nd example}

Let us consider now
\begin{equation}
\mathcal{F}_2 \equiv F_4 \left(\left.\begin{array}{l}1,\,1-\varepsilon\\1-\varepsilon,\,1+\varepsilon\end{array}\right|x_1,\,x_2\right)  \label{iii7}
\end{equation}
From the series expansion
\begin{eqnarray}
\mathcal{F}_2 &=&
\sum_{m_1=0}^\infty \sum_{m_2=0}^\infty \frac{(1)_{m_1+m_2}\,(1-\varepsilon)_{m_1+m_2}}{(1-\varepsilon)_{m_1}\,(1+\varepsilon)_{m_2}}\,
\frac{x_1^{m_1}}{m_1!}\,\frac{x_2^{m_2}}{m_2!}  \nonumber  \\
&=&\sum_{m=0}^\infty \,\sum_{n=0}^m\, m!\,\frac{(1+m-n-\varepsilon)_n}{(1+\varepsilon)_n}\,\frac{x_1^{m-n}}{(m-n)!}\,\frac{x_2^n}{n!}\,.
 \label{iii8}
\end{eqnarray}
we obtain, proceeding as before,
\begin{eqnarray}
\mathcal{F}_2 &=& \sum_{k=0}^\infty \,\varepsilon^k \,(-1)^k\,\sum_{m=0}^\infty \, \,\sum_{n=0}^m\,\left(\delta_{k,\,0}\,(-1)^n - \sum_{j=1}^n(-1)^{j}{m+j \choose n}{n \choose j}\,\frac{1}{j^{k}}\right)  \nonumber  \\
& & \hspace{120pt}\times\,{m \choose n}\,x_1^{m-n}\,x_2^n\,.  \label{iii9}
\end{eqnarray}
The relation (see Eq. 47 of Sec. 4.2.5 in Ref. 27)
\begin{equation}
(-1)^n -\sum_{j=1}^n(-1)^{j}\,{m+j \choose n}{n \choose j} = {m \choose n}  \label{iii10}
\end{equation}
allows one to realize that the coefficient of $\varepsilon^0$ reduces to the same expression as in the preceding example, namely the right hand side of Eq. (\ref{iii6}).

\subsection{3rd example}

The Appell function
\begin{equation}
\mathcal{F}_3 \equiv F_4 \left( \left.\begin{array}{l}1,\,1-\varepsilon\\1+\varepsilon,\,1-\varepsilon\end{array}\right|x_1,\,x_2\right)  \label{iii11}
\end{equation}
can be treated as that in the preceding example by simply interchanging the roles of $x_1$ and $x_2$. Then,
\begin{eqnarray}
\mathcal{F}_3 &=& \sum_{k=0}^\infty \,\varepsilon^k \,(-1)^k\,\sum_{m=0}^\infty \, \,\sum_{n=0}^m\,\left(\delta_{k,\,0}\,(-1)^n - \sum_{j=1}^n(-1)^{j}{m+j \choose n}{n \choose j}\frac{1}{j^{k}}\right)  \nonumber  \\
& & \hspace{120pt}\times\,{m \choose n}\,x_1^n\,x_2^{m-n}\,.  \label{iii12}
\end{eqnarray}

\subsection{4th example}

The case of
\begin{equation}
\mathcal{F}_4 \equiv F_4 \left( \left.\begin{array}{l}1,\,1+\varepsilon\\1+\varepsilon,\,1+\varepsilon\end{array}\right|x_1,\,x_2\right)  \label{iii13}
\end{equation}
has been worked out by Del Duca {\it et al.}, in their above mentioned paper, by using the algebra properties of nested harmonic sums. Our procedure starts with the series expansion
\begin{eqnarray}
\mathcal{F}_4 &=&
\sum_{m_1=0}^\infty \sum_{m_2=0}^\infty \frac{(1)_{m_1+m_2}\,(1+\varepsilon)_{m_1+m_2}}{(1+\varepsilon)_{m_1}\,(1+\varepsilon)_{m_2}}\,
\frac{x_1^{m_1}}{m_1!}\,\frac{x_2^{m_2}}{m_2!}  \nonumber  \\
&=& \sum_{m=0}^\infty \,\sum_{n=0}^m\, m!\,\frac{(1+m-n+\varepsilon)_n}{(1+\varepsilon)_n}\,\frac{x_1^n}{n!}\,\frac{x_2^{m-n}}{(m-n)!}\,.
 \label{iii14}
\end{eqnarray}
to obtain
\begin{eqnarray}
\mathcal{F}_4 &=& \sum_{k=0}^\infty \,\varepsilon^k \,(-1)^k\,\sum_{m=0}^\infty \, \sum_{n=0}^m\,\left(\delta_{k,\,0} - \sum_{j=1}^n(-1)^{j}{m-j \choose n}{n \choose j}\frac{1}{j^{k}}\right)  \nonumber  \\
& & \hspace{120pt}\times\,{m \choose n}\,x_1^{n}\,x_2^{m-n}\,.  \label{iii15}
\end{eqnarray}
Analogously to what occurred in the 1st example, the right hand side of Eq. (\ref{iii15}) is invariant in the interchange $x_1\Longleftrightarrow x_2$. For a better comparison with the results given in Ref. 18, our last equation may be written in the form
\begin{eqnarray}
\mathcal{F}_4 &=& F_4 \left( \left.\begin{array}{l}1,\,1\\1,\,1\end{array}\right|x_1,\,x_2\right) + \sum_{k=1}^\infty \,\varepsilon^k \,(-1)^{k+1}\,\sum_{n_1=0}^\infty \, \sum_{n_2=0}^\infty\,{n_1+n_2 \choose n_1}^2 \,x_1^{n_1}\,x_2^{n_2}  \nonumber  \\
& & \hspace{20pt}\times\,\left(\sum_{j=1}^{n_1}(-1)^{j}\frac{(n_1+1-j)_j\,(n_2+1-j)_j}{(n_1+n_2+1-j)_j\,j!}\,\frac{1}{j^{k}}\right) \,.  \label{iii16}
\end{eqnarray}
As it can be seen, our method provides much simpler results.

\subsection{5th example}

As a last example of Appell function, let us consider
\begin{equation}
\mathcal{F}_5 \equiv F_2 \left(1 \left|\begin{array}{l}\frac{1}{2},\,\frac{3}{2}-\varepsilon\\2-\varepsilon,\,3-2\varepsilon\end{array}\right|\sqrt{z}\,R,\,R\right)  \label{iii17}
\end{equation}
We have for its series expansion
\begin{eqnarray}
\mathcal{F}_5 &=&
\sum_{m_1=0}^\infty \sum_{m_2=0}^\infty(1)_{m_1+m_2}\, \frac{(1/2)_{m_1}\,(3/2-\varepsilon)_{m_2}}{(2-\varepsilon)_{m_1}\,(3-2\varepsilon)_{m_2}}\,
\frac{z^{m_1/2}\,R^{m_1}}{m_1!}\,\frac{R^{m_2}}{m_2!}  \nonumber  \\
&=& \sum_{m=0}^\infty \,R^m\,\sum_{n=0}^m\, {m \choose n} \,(1/2)_n\,\frac{1}{(2-\varepsilon)_n}\,\frac{(3/2-\varepsilon)_{m-n}}{(3-2\varepsilon)_{m-n}}\,z^{n/2}\,.
 \label{iii18}
\end{eqnarray}
Now, it is immediate to write the successive derivatives with respect to $\varepsilon$ in terms of the $\mathcal{Q}_n^{(k)}(2-\varepsilon)$ and the $\mathcal{R}_{m-n,m-n}^{(k)}(3/2-\varepsilon,\,3-2\varepsilon)$ as given in Eqs. (\ref{b3}) and (\ref{ii14}). One obtains in this way for the $\varepsilon$-expansion (assuming that a sum is void if the lower limit of the summation index is larger than the upper one)
\begin{eqnarray}
\mathcal{F}_5 &=& \sum_{k=0}^\infty \,\varepsilon^k \,\sum_{m=0}^\infty  \,R^m\sum_{n=0}^m\, {m \choose n} \,(1/2)_n\,z^{n/2} \nonumber  \\
& & \times\,\sum_{k_1=0}^k \left(\delta_{n,0}\,\delta_{k_1,0}+\sum_{l=0}^{n-1}\frac{(-1)^l}{l!\,(n\! -\! 1\! -\! l)!}\,\frac{1}{(2\! +\! l)^{k_1+1}}\right)  \nonumber  \\
& & \times\,\left(\delta_{k,k_1}\,2^{n-m} + \, \sum_{j=0}^{m-n-1}\frac{(-1)^{j}\,(-j/2)_{m-n}}{j!\,(m\! -\! n\! -\! 1\! -\! j)!}\, \frac{2^{k-k_1}}{(3+j)^{k-k_1+1}}\right)  \nonumber \\
&=& \sum_{k=0}^\infty \,\varepsilon^k \,\sum_{m=0}^\infty  \,\left(\frac{R}{2}\right)^m\,\sum_{n=0}^m\, {m \choose n} \,\Big(\delta_{n,0}+\theta(n-1)\,(2n\! -\! 1)!!\Big)\,z^{n/2} \nonumber \\
& & \times\,\sum_{k_1=0}^k\, \left(\delta_{n,0}\,\delta_{k_1,0}-\sum_{l=1}^{n}\frac{(-1)^{l}}{l!\,(n-l)!}\,\frac{l}{(1+l)^{k_1+1}}\right) \nonumber \\
& &\times\left(\delta_{k,k_1} - \sum_{j=1}^{[(m-n)/2]}\frac{(-1)^{j}\,(2m\! -\! 2n\! -\! 2j\! -\! 1)!!}{2^j\,(m\! -\! n\! -\! 2j)!\,(j\! -\! 1)!}\, \frac{1}{(1\! +\! j)^{k-k_1+1} }  \right).  \label{iii19}
\end{eqnarray}
The last expression can be written in the form
\begin{equation}
\mathcal{F}_5 = \sum_{k=0}^\infty \,\varepsilon^k \,\sum_{m=0}^\infty\,R^m\,\sum_{n=0}^m\, c_{m,n}^{(k)}\, z^{n/2}\,,  \label{iii20}
\end{equation}
with an appropriate definition of the coefficients $c_{m,n}^{(k)}$. We report, in Tables I to IV, these coefficients for the lowest values of  $k$, $m$, and $n$.

\begin{table}
\centering
\caption{First coefficients $c_{m,n}^{(0)}$ of the $\varepsilon$-expansion  Eq. (\ref{iii20}).}
\begin{tabular}{|c|cccccc|}
\hline
\  & $n=0$ & $n=1$ & $n=2$ & $n=3$ & $n=4$ & $n=5$  \\
\hline
$m=0$  & 1 & & & & &  \\
$m=1$  & $\frac{1}{2}$ & $\frac{1}{4}$ &  & & & \\
$m=2$  & $\frac{5}{16}$ & $\frac{1}{4}$ & $\frac{1}{8}$ & & & \\
$m=3$  & $\frac{7}{32}$ & $\frac{15}{64}$ & $\frac{3}{16}$ & $\frac{5}{64}$ & & \\
$m=4$  & $\frac{21}{128}$ & $\frac{7}{32}$ & $\frac{15}{64}$ & $\frac{5}{32}$ & $\frac{7}{128}$ & \\
$m=5$  & $\frac{33}{256}$ & $\frac{105}{512}$ & $\frac{35}{128}$ & $\frac{125}{512}$ & $\frac{35}{256}$ & $\frac{21}{512}$ \\
\hline
\end{tabular}
\end{table}

\begin{table}
\centering
\caption{First coefficients $c_{m,n}^{(1)}$ of the $\varepsilon$-expansion  Eq. (\ref{iii20}).}
\begin{tabular}{|c|cccccc|}
\hline
\  & $n=0$ & $n=1$ & $n=2$ & $n=3$ & $n=4$ & $n=5$  \\
\hline
$m=0$  & 0 & & & & &  \\
$m=1$  & 0 & $\frac{1}{8}$ &  & & & \\
$m=2$  & $\frac{1}{32}$ & $\frac{1}{8}$ & $\frac{5}{48}$ & & & \\
$m=3$  & $\frac{3}{64}$ & $\frac{9}{64}$ & $\frac{5}{32}$ & $\frac{65}{768}$ & & \\
$m=4$  & $\frac{41}{768}$ & $\frac{5}{32}$ & $\frac{7}{32}$ & $\frac{65}{384}$ & $\frac{539}{7680}$ & \\
$m=5$  & $\frac{85}{1536}$ & $\frac{65}{384}$ & $\frac{55}{192}$ & $\frac{1775}{6144}$ & $\frac{539}{3072}$ & $\frac{609}{10240}$ \\
\hline
\end{tabular}
\end{table}

\begin{table}
\centering
\caption{First coefficients $c_{m,n}^{(2)}$ of the $\varepsilon$-expansion  Eq. (\ref{iii20}).}
\begin{tabular}{|c|cccccc|}
\hline
\  & $n=0$ & $n=1$ & $n=2$ & $n=3$ & $n=4$ & $n=5$  \\
\hline
$m=0$  & 0 & & & & &  \\
$m=1$  & 0 & $\frac{1}{16}$ &  & & & \\
$m=2$  & $\frac{1}{64}$ & $\frac{1}{16}$ & $\frac{19}{288}$ & & & \\
$m=3$  & $\frac{3}{128}$ & $\frac{21}{256}$ & $\frac{19}{192}$ & $\frac{575}{9216}$ & & \\
$m=4$  & $\frac{127}{4608}$ & $\frac{13}{128}$ & $\frac{119}{768}$ & $\frac{575}{4608}$ & $\frac{26593}{460800}$ & \\
$m=5$  & $\frac{275}{9216}$ & $\frac{2195}{18432}$ & $\frac{1025}{4608}$ & $\frac{17225}{73728}$ & $\frac{26593}{184320}$ & $\frac{32683}{614400}$ \\
\hline
\end{tabular}
\end{table}

\begin{table}
\centering
\caption{First coefficients $c_{m,n}^{(3)}$ of the $\varepsilon$-expansion  Eq. (\ref{iii20}).}
\begin{tabular}{|c|cccccc|}
\hline
\  & $n=0$ & $n=1$ & $n=2$ & $n=3$ & $n=4$ & $n=5$  \\
\hline
$m=0$  & 0 & & & & &  \\
$m=1$  & 0 & $\frac{1}{32}$ &  & & & \\
$m=2$  & $\frac{1}{128}$ & $\frac{1}{32}$ & $\frac{65}{1728}$ & & & \\
$m=3$  & $\frac{3}{256}$ & $\frac{3}{64}$ & $\frac{65}{1152}$ & $\frac{4325}{110592}$ & & \\
$m=4$  & $\frac{389}{27648}$ & $\frac{1}{16}$ & $\frac{227}{2304}$ & $\frac{4325}{55296}$ & $\frac{1075991}{27648000}$ & \\
$m=5$  & $\frac{865}{55296}$ & $\frac{4265}{55296}$ & $\frac{2105}{13824}$ & $\frac{142475}{884736}$ & $\frac{1075991}{11059200}$ & $\frac{155869}{4096000}$ \\
\hline
\end{tabular}
\end{table}

\section{$\varepsilon$-expansion of Kamp\'e de F\'eriet functions}

Examples of  Kamp\'e de F\'eriet functions can be found also in the repeatedly mentioned paper by Del Duca {\it et al.} (see Eq. (5.13) of Ref. 18),
\begin{eqnarray}
\mathcal{F}_6 & \equiv & F_{0,2}^{2,1} \left(\begin{array}{l}1+\delta\,,\,1+\delta-\varepsilon \\ \,--\,,\,-- \end{array}\left|\begin{array}{l} 1\,,\,--\,,\,--\,,\,-- \\1+\delta\,,\,1-\varepsilon\,,\,1+\varepsilon+\delta\,,\,--\end{array}\right|x_1,\,x_2\right),  \label{iv1}  \\
\mathcal{F}_7 & \equiv & F_{0,2}^{2,1} \left(\begin{array}{l}1+\delta\,,\,1+\delta+\varepsilon \\ \,--\,,\,-- \end{array}\left|\begin{array}{l} 1\,,\,--\,,\,--\,,\,-- \\1+\delta\,,\,1+\varepsilon\,,\,1+\varepsilon+\delta\,,\,--\end{array}\right|x_1,\,x_2\right).  \label{iv2}
\end{eqnarray}
Let us consider the first of them. The series expansion
\begin{eqnarray}
\mathcal{F}_6 &=&\sum_{n_1,n_2=0}^\infty \frac{(1+\delta)_{n_1+n_2}\,(1+\delta-\varepsilon)_{n_1+n_2}\,(1)_{n_1}}{(1+\delta)_{n_1}\,(1-\varepsilon)_{n_1}\,(1+\delta+\varepsilon)_{n_2}}\, \frac{x_1^{n_1}}{n_1!}\,\frac{x_2^{n_2}}{n_2!}  \nonumber  \\
&=&\sum_{n_1,n_2=0}^\infty (1\! +\! n_1\! +\! \delta)_{n_2}\,\frac{(1\! +\! \delta\! -\! \varepsilon)_{n_1}}{(1\! -\! \varepsilon)_{n_1}}\,\frac{(1\! +\! n_1\! +\! \delta\! -\! \varepsilon)_{n_2}}{(1\! +\! \delta\! +\! \varepsilon)_{n_2}}\, x_1^{n_1}\,\frac{x_2^{n_2}}{n_2!}  \label{iv3}
\end{eqnarray}
can be written in the form
\begin{eqnarray}
\mathcal{F}_6 &=& \sum_{n_1,n_2=0}^\infty (1\! +\! n_1\! +\! \delta)_{n_2}\left(1+\sum_{l=0}^{n_1-1}\frac{(-1)^l\,(\delta-l)_{n_1}}{l!\,(n_1-1-l)!}\,\frac{1}{1+l-\varepsilon}\right)  \nonumber \\
& &\hspace{-20pt} \times\,\left((-1)^{n_2}+\sum_{j=0}^{n_2-1}\frac{(-1)^j\,(2\! +\! n_1\! +\! j\! +2\delta)_{n_2}}{j!\,(n_2-1-j)!}\,\frac{1}{1\! +\! j\! +\! \delta\! +\!  \varepsilon}\right)
\, x_1^{n_1}\,\frac{x_2^{n_2}}{n_2!}\,,  \label{iv4}
\end{eqnarray}
which can be easily derived repeatedly with respect to $\varepsilon$ to obtain the $\varepsilon$-expansion
\begin{eqnarray}
\mathcal{F}_6 &=& \sum_{k=0}^\infty\,\varepsilon^k\,\sum_{n_1,n_2=0}^\infty (1\! +\! n_1\! +\!\delta)_{n_2}\, x_1^{n_1}\,\frac{x_2^{n_2}}{n_2!} \nonumber  \\
& & \times\, \sum_{k_1=0}^k (-1)^{k-k_1}\,\left(\delta_{k_1,0}-\sum_{l=1}^{n_1}\frac{(-1)^l\,(1+\delta-l)_{n_1}}{l!\,(n_1-l)!}\,\frac{1}{l^{k_1}}\right)  \nonumber \\
& &\hspace{-20pt} \times\,\left(\delta_{k_1,k}\,(-1)^{n_2}+\sum_{j=0}^{n_2-1}\frac{(-1)^j\,(2\! +\! n_1\! +\! j\! +2\delta)_{n_2}}{j!\,(n_2-1-j)!}\,\frac{1}{(1\! +\! j\! +\! \delta)^{k-k_1+1}}\right).  \label{iv5}
\end{eqnarray}
An alternative expression of this expansion will be obtained in the next section.

Analogously, from the series expansion of the second of the Kamp\'e de F\'eriet functions mentioned above,
\begin{eqnarray}
\mathcal{F}_7 &=&\sum_{n_1,n_2=0}^\infty \frac{(1+\delta)_{n_1+n_2}\,(1+\delta+\varepsilon)_{n_1+n_2}\,(1)_{n_1}}{(1+\delta)_{n_1}\,(1+\varepsilon)_{n_1}\,(1+\delta+\varepsilon)_{n_2}}\, \frac{x_1^{n_1}}{n_1!}\,\frac{x_2^{n_2}}{n_2!}  \nonumber  \\
&=&\sum_{n_1,n_2=0}^\infty (1\! +\! n_1\! +\! \delta)_{n_2}\,\frac{(1\! +\! n_2\! +\! \delta\! +\! \varepsilon)_{n_1}}{(1\! +\! \varepsilon)_{n_1}}\, x_1^{n_1}\,\frac{x_2^{n_2}}{n_2!}\,,  \label{iv6}
\end{eqnarray}
written as
\begin{eqnarray}
\mathcal{F}_7 &=& \sum_{n_1,n_2=0}^\infty (1\! +\! n_1\! +\! \delta)_{n_2}\, x_1^{n_1}\,\frac{x_2^{n_2}}{n_2!}  \nonumber \\
& &\hspace{30pt} \times\,\left(1+\sum_{j=0}^{n_1-1}\frac{(-1)^j\,(n_2+\delta-j)_{n_1}}{j!\,(n_1-1-j)!}\,\frac{1}{1+j+\varepsilon}\right)\,,  \label{iv7}
\end{eqnarray}
one obtains the $\varepsilon$-expansion
\begin{eqnarray}
\mathcal{F}_7 &=& \sum_{k=0}^\infty\,\varepsilon^k\,(-1)^k\sum_{n_1,n_2=0}^\infty (1\! +\! n_1\! +\!\delta)_{n_2}\, x_1^{n_1}\,\frac{x_2^{n_2}}{n_2!} \nonumber  \\
& & \hspace{30pt}\times\,\left(\delta_{k,0}-\sum_{j=1}^{n_1}\frac{(-1)^j\,(n_2+1+\delta-j)_{n_1}}{j!\,(n_1-j)!}\,\frac{1}{j^{k}}\right).  \label{iv8}
\end{eqnarray}

In Eq. (5.13) of Ref. 18, $\mathcal{F}_6$ and $\mathcal{F}_7$ do not appear as such, but through their first derivative with respect to $\delta$ taken at $\delta=0$,
\[
\left.\frac{\partial}{\partial\delta}\mathcal{F}_6\right|_{\delta=0}\,, \qquad \left.\frac{\partial}{\partial\delta}\mathcal{F}_7\right|_{\delta=0}\,.
\]
There is no difficulty to calculate the first derivatives with respect to $\delta$ of the right hand sides of Eqs. (\ref{iv5}) and (\ref{iv8}). Let us consider, for instance, the last of these. We obtain
\begin{eqnarray}
\left.\frac{\partial}{\partial\delta}\mathcal{F}_7\right|_{\delta=0} &=& \sum_{k=0}^\infty\,\varepsilon^k\,(-1)^k\sum_{n_1,n_2=0}^\infty
x_1^{n_1}\,\frac{x_2^{n_2}}{n_2!}  \nonumber  \\
& & \hspace{20pt} \times\,\left\{ \mathcal{P}_{n_2}^{(1)}(1\! +\! n_1)\left(\delta_{k,0}-\sum_{j=1}^{n_1}\frac{(n_2\! +\! 1-\! j)_{n_1}}{j!\,(n_1\! -\! j)!}\,\frac{(-1)^j}{j^{k}}\right) \right.\nonumber  \\
& &\hspace{40pt}\left. -\,\frac{(n_1\! +\! n_2)!}{n_1!}\left(\sum_{j=1}^{n_1}\frac{\mathcal{P}_{n_1}^{(1)}(n_2\! +\! 1\! -\! j)}{j!\,(n_1\! -\! j)!}\,\frac{(-1)^j}{j^{k}}\right)\right\},  \label{iv9}
\end{eqnarray}
which can be reduced, with the aid of Eq. (\ref{A19}), to
\begin{eqnarray}
\left.\frac{\partial}{\partial\delta}\mathcal{F}_7\right|_{\delta=0} &=& \sum_{k=0}^\infty\,\varepsilon^k\,(-1)^k\sum_{n_1,n_2=0}^\infty
x_1^{n_1}\,\frac{x_2^{n_2}}{n_2!}  \nonumber  \\
& &  \times\left\{ (-1)^{n_2-1}\,n_2\, B_{n_2-1}^{(n_2+1)}(-n_1)\left(\delta_{k,0}-\sum_{j=1}^{n_1}
\frac{(n_2\! +\! 1\! -\! j)_{n_1}}{j!\,(n_1\! -\! j)!}\,\frac{(-1)^j}{j^{k}}\right) \right.\nonumber  \\
& & \left. +\,(-1)^{n_1}\,n_1\,\frac{(n_1\! +\! n_2)!}{(n_1!)^2}\,\sum_{j=1}^{n_1}\,{n_1 \choose j}\, \frac{(-1)^j}{j^{k}}\,B_{n_1-1}^{(n_1+1)}(-n_2\! +\!j) \right\}.\label{iv10}
\end{eqnarray}

\section{Partial fraction decomposition of more general quotients of Poch\-hammer symbols}

The partial fraction decomposition considered in Section IV can be applied to the quotient of products of several Pochhammer symbols. The extension is immediate if all roots of the denominator, considered as a function of $\varepsilon$, are simple. From Eqs. (\ref{ii1}) and (\ref{ii4}), with the notation introduced in Eq. (\ref{i1}), one can obtain easily
\begin{eqnarray}
\frac{\prod_{p=1}^r (\alpha_p)_{m_p}}{\prod_{q=1}^s(\beta_q)_{n_q}}&=&\sum_{q=1}^s\,\sum_{j_q=0}^{n_q-1}\frac{C_{q,j_q}}{B_q\! +\! j_q\! +\! b_q\,\varepsilon}\,, \label{v1}  \\
& & \hspace{120pt} \mbox{for}\qquad \sum_{p=1}^r m_p < \sum_{q=1}^s n_q\,, \nonumber
\end{eqnarray}
\begin{eqnarray}
\frac{\prod_{p=1}^r (\alpha_p)_{m_p}}{\prod_{q=1}^s (\beta_q)_{n_q}}&=&\frac{a_1^{m_1}a_2^{m_2}\cdots a_r^{m_r}}{b_1^{n_1}b_2^{n_2}\cdots b_s^{n_s}} + \sum_{q=1}^s\,\sum_{j_q=0}^{n_q-1}\frac{C_{q,j_q}}{B_q\! +\! j_q\! +\! b_q\,\varepsilon}\,, \label{v2} \\
& & \hspace{120pt} \mbox{for}\qquad \sum_{p=1}^r m_p =  \sum_{q=1}^s n_q\,,    \nonumber
\end{eqnarray}
with coefficients
\begin{equation}
C_{q,j_q}=\frac{(-1)^{j_q}}{j_q!\,(n_q\! -\! 1\! -\! j_q)!}\,\frac{\prod_{p=1}^r\Big(A_p-(a_p/b_q)(B_q\! +\! j_q)\Big)_{m_p}} {\prod_{k=1, k\neq q}^s \Big(B_k-(b_k/b_q)(B_q\! +\! j_q)\Big)_{n_k}}\,.   \label{v3}
\end{equation}

As an application of these partial fraction decompositions, let us obtain an expression of the $\varepsilon$-expansion of $\mathcal{F}_6$ alternative to that given in Eq. (\ref{iv5}). In view of Eqs. (\ref{v2}) and (\ref{v3}), we have
\begin{eqnarray}
\frac{(1\! +\! \delta\! -\! \varepsilon)_{n_1+n_2}}{(1\! -\! \varepsilon)_{n_1}\,(1\! +\! \delta\! +\! \varepsilon)_{n_2}} &=& (-1)^{n_2}\! + \! \sum_{j_1=0}^{n_1-1} \frac{(\delta\! -\! j_1)_{n_1+n_2}}{(2\! +\! \delta\! +\! j_1)_{n_2}}\,\frac{(-1)^{j_1}}{j_1!\,(n_1\! -\! 1\! -\! j_1)!}\,\frac{1}{1\! +\! j_1\! -\! \varepsilon} \nonumber  \\
& & \hspace{-40pt}+\,\sum_{j_2=0}^{n_2-1}\, \frac{(2\! +\! 2\delta\! +\! j_2)_{n_1+n_2}}{(2\! +\! \delta\! +\! j_2)_{n_1}}\,\frac{(-1)^{j_2}}{j_2!\,(n_2\! -\! 1\! -\! j_2)!}\,\frac{1}{1+\! j_2\! +\! \delta\! +\! \varepsilon}\,.  \label{v4}
\end{eqnarray}
By substitution of this expression in the first of Eqs. (\ref{iv3}), repeated derivation with respect to $\varepsilon$, and particularization for $\varepsilon=0$, one obtains the expansion
\begin{eqnarray}
\mathcal{F}_6 &=& \sum_{k=0}^\infty\,\varepsilon^k\,\sum_{n_1,n_2=0}^\infty (1\! +\! n_1\! +\!\delta)_{n_2}\, x_1^{n_1}\,\frac{x_2^{n_2}}{n_2!} \nonumber  \\
& & \times\, \left\{\delta_{k,0}\,(-1)^{n_2} - \sum_{j_1=1}^{n_1} \frac{(1\! +\! \delta\! -\! j_1)_{n_1+n_2}}{(1\! +\! \delta\! +\! j_1)_{n_2}}\,\frac{(-1)^{j_1}}{j_1!\,(n_1\! -\! j_1)!}\,\frac{1}{j_1^k}\right. \nonumber  \\
& & \left.+\,(-1)^k\,\sum_{j_2=0}^{n_2-1} \frac{(2\! +\! 2\delta\! +\! j_2)_{n_1+n_2}}{(2\! +\! \delta\! +\! j_2)_{n_1}}\,\frac{(-1)^{j_2}}{j_2!\,(n_2\! -\! 1\! -\! j_2)!}\,\frac{1}{(1\! +\! j_2\! +\! \delta)^{k+1}}\right\}.  \label{v5}
\end{eqnarray}

\section*{Acknowledgments}

The idea of this work arose from a private communication of Oleg V. Tarasov. The presentation of our results has benefited considerably from the comments of an anonymous reviewer. In particular, the generating relation given in Eq. (\ref{nueva1}) has been suggested by the reviewer. We acknowledge financial support of the EU under Contract MTRN-CT-2006-035482 (FLAVIAnet) and by MUIR, Italy, under Project 2005-023102, of Gobierno de Arag\'on and Fondo Social Europeo (Project E24/1) and Ministerio de Ciencia e Innovaci\'on (Project MTM2009-11154), and of the Spanish Government and ERDF funds from the EU Commission [grants FPA2007-60323, CSD2007-00042 (Consolider Project CPAN)].

\end{document}